\documentclass{aa}  

\usepackage{graphicx}
\usepackage{txfonts}
\usepackage{arydshln}
\usepackage{natbib}
\bibpunct{(}{)}{;}{a}{}{,}
\begin{document}

   \title{The pre-eruption state of T CrB as observed with ALMA in 2024}

   \author{D. Petry\inst{1}
          \and
          G. Sala\inst{2,3}
          \and
          I. El Mellah\inst{4,5}
          \and
          T. Stanke\inst{6}
          \and
          J. Greiner\inst{6}
          }

   \institute{
             European Southern Observatory, Karl-Schwarzschild-Strasse 2, 85748 Garching, Germany \\ 
             \email{dpetry@eso.org}
        \and
        Departament de Física, EEBE, Universitat Politècnica de Catalunya, c/Eduard Maristany 16, 08019 Barcelona, Spain \\
              \email{gloria.sala@upc.edu}
         \and
         Institut d'Estudis Espacials de Catalunya (IEEC), 08860 Castelldefels (Barcelona), Spain 
         \and
        Departamento de Física, Universidad de Santiago de Chile, Av. Victor Jara 3659, Santiago, Chile
        \and
        Center for Interdisciplinary Research in Astrophysics and Space Exploration (CIRAS), USACH, Santiago, Chile
        \and
             Max-Planck Institut für extraterrestrische Physik (MPE), Giessenbachstr. 1, 85748 Garching, Germany
            }
   \date{5 June 2025 ; 5 September 2025}
 
  \abstract
   {T~CrB is a nearby symbiotic binary and a recurrent nova with a period of ca. 80 years. The next eruption is expected to take place in 2025 or 2026. As part of a global multi-wavelength campaign on the event, we have obtained
   time on the Atacama Large mm/sub-mm Array to observe the object between 42~GHz and 407~GHz.}
   {In this first paper on our results, we present our pre-eruption observations made in ALMA frequency Bands~1, 3, 4, 6, 7, and 8 in August to November 2024 and constrain the properties of the environment into which the imminent next nova will erupt.}
   {We calibrate and image our ALMA data following the standard ALMA procedures, search for line emission,
   and construct a spectrum 
   from the points for orbital phase 0.43 (August 2024) from 44~GHz to 350~GHz.
   We compare this with the spectra we measure in the VLA data obtained by Linford et al. in 2016/17 over the upper half of their frequency range (13.5~GHz to 35~GHz). 
   We create aggregate bandwidth images from all our 2024 data and, for maximum angular resolution, from the band 7 and 8 data from which we compute an upper limit on the brightness temperature in an annulus with radius 0.8~arcsec - 1.6~arcsec.}
   {In the second half of 2024, after the end of its latest high state, 
   the quiescent T~CrB was a faint mm source with a spectral energy distribution well described by a powerlaw with
   index $\alpha=$0.56$\pm$0.11 and a flux density of ca. 0.1~mJy at 44~GHz and 0.4~mJy at 400~GHz.
   There is no significant line emission.
   This is in agreement with expectations for free-free emission from the partially ionized wind of
   the red giant donor star and, in extrapolation to 35~GHz, a factor 5 fainter than the emission observed in 2016/17 during the latest high state.
   Comparing the spectra from that high-state between 13.5~GHz and 35~GHz with our spectrum from 2024, our spectrum is softer. The spectral index is on average lower by 0.34$\pm$0.11 .
   Our per-band and aggregate bandwidth images of T~CrB show an unresolved point source with no evidence for extended structure. 
   }
   {A simple model of a free-free emitting, fully-ionized stellar wind seems to describe well the 2016/17 high state of T~CrB but not our 2024 ALMA measurements with their low flux and high turnover frequency suggesting that in 2024, the wind was far from fully ionized.}

   \keywords{binaries: symbiotic --
                novae, cataclysmic variables --
                stars: individual: T~CrB
               }

   \maketitle
%

\section{Introduction}

Nova outbursts are the result of thermonuclear runaway in the hydrogen-rich material accumulated on the surface of a white dwarf (WD) in an accreting binary system \citep{2016PASP..128e1001S}. Unlike Type Ia supernovae, nova eruptions do not destroy the WD. The material expelled by the nova outburst, enriched with newly formed elements, ranges from $10^{-7}$ to $10^{-4} M_{\odot}$ and is propelled at velocities of several thousand km/s \citep{2016PASP..128e1001S, 1998ApJ...494..680J}. Novae are events which typically recur on timescales of $10^4$ to $10^5$ years. A particularly extreme and interesting subclass of novae are "recurrent novae" (RNe) characterized by multiple recorded eruptions with recurrence intervals between 1 and 100 years.

Novae are relatively frequent in our Galaxy with an estimated occurrence rate of 50($^{+31}_{-23}$) per year \citep{2017ApJ...834..196S}. However, only about 20\% are actually observed, as many are obscured by interstellar dust. Among the more than 500 novae known in the Galaxy to date, only 10 recurrent novae are confirmed \citep{2021gacv.workE..44D}. The short recurrence times suggest that these WDs are near the Chandrasekhar mass limit and/or are accreting at high rates. Hydrodynamical simulations show that the mass of the ejected envelope in nova outbursts on massive white dwarfs is smaller and less mixed with the WD core than in the case of smaller WD masses \citep{2008ASPC..401..131T, 2016stex.book.....J}. This supports the scenario in which massive WDs in RNe accumulate mass over time, making them promising progenitor candidates for Type Ia supernovae \citep{2012BASI...40..393K}.

Approximately half of the known RNe are hosted in binary systems with red giant donors and long orbital periods, forming a subclass known as Symbiotic-like Recurrent Novae (SyRNe).  While the explosion mechanism remains consistent with that of classical novae, the crucial difference is the presence of dense circumstellar material surrounding the WD \citep{1993ApJ...416..355S, 2022RNAAS...6...92A, 2023A&A...674A.139A} which stems mostly from the red giant wind. In the initial weeks following the outburst, the interaction between the nova ejecta and the red giant wind generates a radiative shock that drives an ionizing precursor through the wind. The advance of this ionization front can be traced via changes in the UV and optical absorption spectra, as well as by narrow recombination emission lines—resembling the narrow line regions seen in AGN. The WD mass sets both the kinetic power and expansion speed of the ejecta. For a detailed discussion see, e.g., \cite{2025CoSka..55c..47M}.

The best-known example of the SyRN class is RS Oph, which undergoes outbursts approximately every 15 years. The most recent events in 1985, 2006, and 2021 were extensively observed across the electromagnetic spectrum, from radio to very high energy (VHE) gamma-rays exceeding 100~GeV \citep{2006Natur.442..276S, 2008ASPC..401...19S}. Up to now, RS~Oph remains the only nova detected at TeV energies \citep{2022NatAs...6..689A} despite attempts to detect VHE emission for other novae \citep{2015ICRC...34..731L}. RS~Oph also was the first RN to be detected in the radio band during its 1985 outburst, exhibiting an expansion rate of the ejecta of 1.5 mas/day at 15 GHz \citep{1986ApJ...305L..71H}.

During the 2006 and 2021 eruptions, the nova ejecta evolved into a rapidly expanding "mini-supernova remnant", featuring synchrotron-emitting flows collimated into jet-like structures \citep{2006Natur.442..279O, 2008ApJ...685L.137S, 2008ApJ...688..559R, 2024MNRAS.528.5528N}. Three-dimensional simulations tracking the system from the accretion phase through to the explosion reproduce the asymmetric geometry arising from binary interactions, and constrain the physical conditions as the ejecta expand into the dense stellar wind of the red giant companion \citep{2008A&A...484L...9W, 2009A&A...493.1049O}. In these outbursts, the expanding shell reached velocities of approximately 5000 km/s, with an estimated ejected mass of $\sim$10$^{-6}$ M$_\odot$. Evidence for bipolar lobes and a density enhancement on the orbital plane was found thanks to the European VLBI Network observations during the 2021 outburst \citep{2022A&A...666L...6M}, while the presence of synchrotron jets was ruled-out using MeerKAT/LOFAR low-frequency radio observations of the same outburst \citep{2023MNRAS.523..132D}.

T~CrB has a recurrence time of $\sim$80~years \citep[last recorded eruptions in 1866 V$\sim$2~mag, 1946 V$\sim$3~mag;][]{2023JHA....54..436S}, 
 and as a donor an M4III giant \citep{1999A&AS..137..473M}. T~CrB is approximately three times closer to the Sun than RS Oph and the nearest known recurrent nova. 
Its exact distance and stellar masses, however, are still under debate since the most recent estimate from Gaia data \citep[896$\pm23$~pc;][]{2021AJ....161..147B} seems to be in conflict with the most 
up-to-date orbital fits which give a significantly smaller distance of 
$752\pm0.4$~pc \citep{2025ApJ...983...76H}. \cite{2025ApJ...983...76H} determine the WD and the red giant companion to have masses of 1.37~M$_\odot$ and 0.69~M$_\odot$, respectively, assuming the Gaia distance. However, the same authors explore the results obtained for the masses with the distance as a free fit parameter and arrive at the smaller distance mentioned above, and also obtain smaller masses: $1.31\pm0.05$~M$_\odot$ for the WD and $0.419^{+0.002}_{-0.004}$~M$_\odot$. Independent
of the particular value of the distance considered, the binary orbit is found to be circular to a high precision with period of 227.55~days.
 
After a high state which started in 2015, T~CrB went into a pre-eruption dip in Mar 2023, indicating an eruption at any time before 2027, more likely in 2025 \citep{2023MNRAS.524.3146S} with the potential to become the brightest naked-eye nova in our life-times \citep{2025ApJ...982...89S}. All this is prompting the community to embark on an extensive
coordinated effort to follow up this event. Because of its closeness, brightness, and predictability – its rare, but expected eruption offers a unique opportunity to prepare an extensive campaign to study the nova phenomenon at all wavelengths. In this context, we present here pre-outburst ALMA observations of T~CrB, aimed at characterizing the pre-eruption state of the object and the medium into which the nova ejecta will expand. 

Section~\ref{sec-obs} describes our ALMA data set of pre-burst observations while
section~\ref{sec-cal} describes the general calibration and imaging procedures.
In section~\ref{sec-spec} we construct a pre-burst spectrum of T~CrB and search for
evidence of extended structure. We then compare in section~\ref{sec-vlaspec} our spectrum with the ones we obtain
from the VLA data at 13.5~GHz - 35~GHz by~\cite{2019ApJ...884....8L} for different days
during the T~CrB high state in 2016/17. 
Section \ref{sec-modelfit} discusses a theoretical interpretation of the 2016/17 and 2024 spectra based on a model of a free-free emitting ionized stellar wind. We summarize our conclusions in section~\ref{sec-conc}.

\section{Observations}
\label{sec-obs}
Table \ref{table:obs} summarizes the ALMA observations we obtained between 10 August and 22 November 2024 in order to characterize the pre-burst state of T~CrB. The object was planned to be observed with the ALMA main array (12 m antennas) in frequency bands 1, 3, 4, 6, 7, and 8 in a near-concurrent
fashion such that a wide-band snap-shot at a single binary orbital phase
could be constructed. On 10 August 2024 (JD 2460533), it was possible to obtain 
near-concurrent observations of bands 1, 3, 4, 6, and 7.
Among these, the band 1 observation (marked "1a" in column "ALMA Band") was taken 
with 180$^\circ$~Walsh switching turned off, which can in principle lead to increased
noise and excessive loss of sensitivity due to automated removal (flagging) of data
by the ALMA online system. However, in this observation no such effect was observed.
The data were scrutinized, passed all quality control tests, and were thus kept in the
data set.

For reasons of inadequate weather conditions, the highest band observations (originally planned
as band 9 but later switched to band 8) could only be obtained 26 days later.
Meanwhile the analysis of the data from 10 August had revealed the unexpected faintness
of the object and supplemental observations were requested in bands 1 and 7 to increase
sensitivity. These were obtained 36 days (band 7) and 100 days (band 1) after 10 August.
They are marked "7b" and "1b" in the column "ALMA Band" of Table~\ref{table:obs}. In the case of the Band "1b" observations, one of the two continuum spectral windows was accidentally placed at a central frequency of 40~GHz instead of 46~GHz which led to an overall shift of the centroid frequency of all combined band "1b" data to 42.089~GHz instead of 44.014~GHz observed for the band "1" data.  
And because of the disabled Walsh switching in our first band 1 observation,
we were granted a re-observation 104 days after 10 August (marked "1" in "ALMA Band").
Our complete pre-burst data thus spans 104 days, a little less than half the object's orbital period.

In preparation of the post-burst observations, also two test observations with the ALMA 7~m array (ACA)
in band 3 and 6 were obtained (same spectral setup as for the main array observation). They are shown in Table \ref{table:obs} with a $^*$ behind the band number.
As the table shows, these observations have much inferior sensitivity and thus did not result in a
detection of T~CrB. They are shown here only for completeness and are not further included in the analysis. 

\begin{table*}[t]
\caption{Summary of the ALMA observations analyzed in this paper.}             
\label{table:obs}      
\centering          
\begin{tabular}{l c r c c c c c c}  
\hline\hline       
ALMA MOUS UID & centroid MJD & time on & number of & ALMA & $\nu_{centroid}$ & agg. BW & beam & RMS\\
              &             & source (s) & antennas & Band & (GHz) & (GHz) & (arcsec) & ($\mu$Jy/beam)\\ 
\hline
uid://A001/X378a/X32 &	60532.887	& 302	& 42 & 3 & 107.804 & 6.54 & 1.21$\times$0.563 & 47 \\
uid://A001/X378a/X3a &	60532.898	& 302	& 42 & 6 & 224.533 & 7.06 & 0.543$\times$0.274 & 52\\
uid://A001/X378a/X36 &	60532.922	& 302	& 41 & 4 & 136.307 & 4.85 & 0.806$\times$0.414 & 38 \\
uid://A001/X378a/X2e &  60532.944 & 362   & 41 & 1a & 44.079 & 5.05 & 2.29$\times$1.34 & 34 \\
uid://A001/X378a/X3e &	60532.969	& 302	& 43 & 7 & 349.507 & 6.88 & 0.303$\times$0.184 & 114\\
\hdashline
uid://A001/X378a/X1d6 &	60559.368	& 3204	& 43 & 8 & 399.993 & 6.58 & 0.517$\times$0.406 & 56\\
uid://A001/X3789/X9e &	60568.775	& 574	& 44 & 7b & 349.507 & 7.07 & 0.808$\times$0.427 & 79\\
uid://A001/X37b5/X117 &	60586.836	& 1118	& 10 & 6$^*$ & 224.379 & 7.27 & 6.82$\times$5.78 & 280\\
uid://A001/X37b5/X104 &	60588.814	& 1239	& 9 & 3$^*$ & 108.209 & 7.2 & 14.2$\times$11.4 & 310 \\
uid://A001/X3789/X9a &	60632.665	& 1330	& 41 & 1b & 42.089 & 5.34 & 8.14$\times$5.68 & 17\\
uid://A001/X378a/X2e &	60636.688	& 362	& 44 & 1 & 44.014 & 5.34 & 7.44$\times$6.12 & 39 \\
\hline                  
\end{tabular}
\tablefoot{
Frequencies are in the  Kinematic Local Standard of Rest (LSRK) reference frame. The dashed line separates the near-concurrent observations obtained on MJD 60532 (10 August 2024) from all later observations. All observations were made with the ALMA 12m array 
except for those marked with "$^*$" on the band number, which were observed with the 7~m array (ACA)
as a test for post-burst observations.
Supplemental 12m array observations, which were
obtained to increase sensitivity after the extreme faintness of the object was noticed, are marked
with a "b" on the band number. In the case of band 1, these observations were obtained
before the original MOUS could be properly observed. Also in the case of band 1, there are two execution blocks available for MOUS uid://A001/X378a/X2e which were taken three months apart
and were analyzed separately.
The first one is marked with an "a" on the band number. See text for details. The noise RMS given in the table refers to the full aggregate bandwidth as given in the column titled "agg. BW".
}
\end{table*}

Each of our ALMA Member Observation Unit Sets (MOUSs) corresponds to a single "Execution Block" (EB) except for the Band 8 observation (which corresponds to two EBs observed on consecutive nights), and as explained above, the Band 1 
MOUS uid://A001/X378a/X2e (which was re-observed and also has two EBs, however, 104 days apart). 

The spectral setup of each MOUS tried to enable the detection
of various potential bright spectral lines while at the same time also achieving good continuum sensitivity. 
As will be described in more detail in the following sections, the object was found in a state
so faint that it was barely possible to detect the object at all in the
individual frequency bands. Each spectral window (SPW) underwent the sophisticated line search
of the ALMA QA Pipeline \citep[version 2024.1.0.8, see][]{2023PASP..135g4501H} and was in addition scrutinized by visual inspection. No line emission was detected.
In table \ref{table:speclimits} we summarize the spectral setup and the
achieved sensitivity per spectral channel.

\begin{table}[t]
\caption{Spectral setup of the ALMA observations analyses in this paper and the achieved sensitivity per spectral channel (1 $\sigma$ noise RMS) with frequencies in the LSRK reference frame and band labels defined as in Table \ref{table:obs}.}             
\label{table:speclimits}      
\centering          
\begin{tabular}{c c c c c c}  
\hline\hline       
Band & SPW & $\nu_{centroid}$ & number of & $\Delta\nu$ & RMS \\ 
     &     & (GHz) & channels & (MHz) & (mJy/beam) \\
\hline                    
1 & 1 & 42.200 & 117 & 15.6239 & 0.36 \\
1 & 2 & 45.950 & 117 & 15.6239 & 0.44 \\
1 & 3 & 43.122 & 957 & 0.0610 & 6.3 \\
1 & 4 & 42.821 & 477 & 0.1221 & 5.7 \\
1 & 5 & 43.424 & 957 & 0.1221 & 5.9 \\
1 & 6 & 44.100 & 1917 & 0.9765 & 2.2 \\
1b & 1 & 39.953 & 117 & 15.6240 & 0.16 \\
1b & 2 & 42.200 & 117 & 15.6240 & 0.17 \\
1b & 3 & 43.122 & 957 & 0.0610 & 3.0 \\
1b & 4 & 42.821 & 478 & 0.1221 & 2.7 \\
1b & 5 & 43.424 & 957 & 0.1221 & 2.8 \\
1b & 6 & 44.099 & 1917 & 0.9765 & 1.0 \\
\hline
3 & 1 & 101.00 & 117 & 15.6252 & 0.38  \\
3 & 2 & 102.80 & 117 & 15.6252 & 0.37 \\
3 & 3 & 113.20 & 1917 & 0.9766 & 3.2 \\
3 & 4 & 114.95 & 1918 & 0.9766 & 4.7\\
\hline
4 & 1 & 131.30 & 117 & 15.6252 & 0.39 \\
4 & 2 & 141.50 & 117 & 15.6252 & 0.36 \\
4 & 3 & 143.30 & 117 & 15.6252 & 0.39 \\
4 & 4 & 129.36 & 478 & 0.1221 & 6.3 \\
4 & 5 & 128.46 & 478 & 0.1221 & 8.2 \\
4 & 6 & 130.27 & 958 & 0.1221 & 6.0 \\
\hline
6 & 1 & 232.70 & 117 & 15.6252 & 0.56 \\
6 & 2 & 216.40 & 1918 & 0.9766 & 3.2 \\
6 & 3 & 219.80 & 1917 & 0.9766 & 4.0 \\
6 & 4 & 230.85 & 1918 & 0.9766 & 3.8 \\
\hline
7 & 1 & 344.50 & 117  & 15.6253 & 1.2 \\
7 & 2 & 342.70 & 117  & 15.6253 & 1.3 \\
7 & 3 & 356.30 & 117  & 15.6253 & 1.6 \\
7 & 4 & 354.51 & 1918 &  0.9766 & 8.1 \\
7b & 1 & 344.50 & 117 & 15.6251 & 0.84 \\
7b & 2 & 342.70 & 117 & 15.6251 & 0.90 \\
7b & 3 & 356.30 & 117 & 15.6251 & 1.1 \\
7b & 4 & 354.51 & 1917 & 0.9766 & 5.5 \\
\hline
8 & 1 & 393.00 & 115 & 15.6256 & 0.85 \\
8 & 2 & 395.00 & 115 & 15.6256 & 0.78 \\
8 & 3 & 405.00 & 115 & 15.6256 & 0.65 \\
8 & 4 & 407.00 & 115 & 15.6256 & 0.63 \\
\hline                  
\end{tabular}
\end{table}

\section{Calibration, imaging, and flux density measurements}
\label{sec-cal}

Each individual dataset was calibrated by the standard ALMA Quality Assurance (QA) Pipeline \citep[version 2024.1.0.8; ][]{2023PASP..135g4501H}  as part of the ALMA QA process.
We screened the quality of the data based on the Pipeline logs delivered
with them and found that no improvement of the official calibration was
possible nor needed. The visibility data was then Fourier transformed and deconvolved ("imaged") using the CASA data
analysis package \citep[CASA 6.6.1;][]{2022PASP..134k4501C} in the following ways:
\begin{enumerate}
\item On a per-day, per-SPW basis (after continuum subtraction) with full spectral resolution to search for line emission. This happened already as part of the ALMA QA processing with the ALMA Pipeline. As mentioned in the previous section, no line emission was detected. As a first indication of the
line sensitivity achieved by our observations, Table \ref{table:speclimits} shows the per-channel noise RMS of all spectral windows. 
\item On a per-band/per-visit (equivalent to per-MOUS) basis over the aggregate bandwidth to maximize continuum sensitivity and obtain flux density measurements (Table \ref{table:fluxes}). From these, a wide-band spectrum can be constructed.
\item All MOUSs and all bands combined with full aggregate bandwidth using Multi-Term Multi-Frequency Synthesis \citep[MTMFS;][]{2011A&A...532A..71R} in order to verify the spectral
parameters of T~CrB obtained from the per-band measurements with an alternative method and search for
potential spatial features in the image (Fig. \ref{fig:image}).
\item Using only bands 7 and 8 to maximize angular resolution (Figs. \ref{fig:trc-b78-cont} and \ref{fig:trc-b7-peak}). See section \ref{sec-spec}.
\end{enumerate}

\begin{table}[t]
\caption{T~CrB flux density measurements for our ALMA main array (12~m) observations described in Table~\ref{table:obs}.}  
\label{table:fluxes}      
\centering          
\begin{tabular}{c c r l}
\hline\hline
centroid MJD & orb. phase & $\nu_{centroid}$ & flux density \\
           &             & (GHz)            & ($\mu$Jy) \\
\hline
60532.887	& 0.4328 & 107.804 & 216$\pm$47$\pm$11 \\
60532.898	& 0.4329 & 224.533 & 251$\pm$52$\pm$13 \\
60532.922	& 0.4329 & 136.307 & 212$\pm$38$\pm$11 \\
60532.944   & 0.4330 &  44.079 & 111$\pm$34$\pm$2.8\\ 
60532.969	& 0.4331 & 349.507 & 435$\pm$114$\pm$22 \\
60559.368	& 0.5492 & 399.993 & 434$\pm$56$\pm$22 \\
60568.775	& 0.5905 & 349.507 & 338$\pm$79$\pm$17 \\
60632.665	& 0.8712 & 42.089  & 181$\pm$17$\pm$4.5 \\
60636.688	& 0.8889 & 44.014  & 182$\pm$39$\pm$4.6 \\
\hline
\end{tabular}
\tablefoot{The orbital phase is computed from the centroid JD following \cite{2000AJ....119.1375F}.
The first errors on the flux density measurements only represent the 1~$\sigma$ statistical errors due to noise. The second errors are the additional systematic uncertainty of the ALMA absolute flux calibration of 2.5\% in bands 1~-~5
and 5\% in bands 6~-~8 (away from strong atmospheric absorption as in our case)\citep{almathbcy11}.}
\end{table}

In order to obtain per-band/per visit images of the target, we produce MTMFS images from the combination
of all target data from one MOUS using the CASA tclean task with standard ALMA QA parameters, in particular
using deconvolver "mtmfs" (with nterms=2 where the fractional bandwidth is $>$10\%), and Briggs weighting with robust=0.5. 
All images show a central emission compatible with a point-source
after deconvolution ("clean")
using a tight elliptical mask around the image center
(the image resembles the point-spread function (PSF) of the interferometer) . 
Few iterations and one major cycle suffice to
obtain a noise-like residual. We obtain the final estimate of the noise RMS from an annulus in the
non-primary-beam-corrected image between ca. 30\% and 60\% of the primary beam radius around the image center.
The flux density of the target is obtained by fitting a Gaussian to an elliptical region of 
the size of the synthesized beam around the peak emission (which we find to be within a synthesized beam radius of the image center, the nominal position of T~CrB\footnote{The T~CrB position used was (ICRS) RA 15:59:30.154, DEC +25.55.12.909 (Epoch 2025-08-10, 22:40:20.4 UT), with proper motion in RA = -6.853E-16 rad/s, in DEC = 1.846E-15 rad/s in agreement with the values from \cite{gaia_edr3}.}). The pixel width was chosen to be 10\% of
the minor axis of the synthesized beam size (see column "beam" in Table \ref{table:obs}).
The flux density is then taken as the peak of the fitted Gaussian.
Table \ref{table:fluxes} shows the resulting flux densities together with the centroid
time (MJD) and frequency of the observation. Also shown is the orbital phase of T~CrB \citep[according to][]{2000AJ....119.1375F} corresponding to the time centroid.
This phase is based on the radial velocity of the M giant in the binary
and defined to be zero when the radial velocity is maximal.

\section{Spectral and spatial analysis}
\label{sec-spec}
The spectrum of T~CrB at mm wavelengths in quiescence, 
long after a thermonuclear explosion has taken place on the WD, is expected
to be time-dependent on scales of days or weeks for two main reasons. First, the emission is a composite
of the emission from near the WD accretion disk, the emission from the red giant, and the emission from its wind (which is mostly free-free emission by the ionized part of the wind).
Due to the rotation of the binary system with orbital period 227.6~days, these three origins of emission are seen in different projections,
and they partially occult each other (the inclination of the system is between $49^\circ$ and $56^\circ$ according to the latest analysis by \cite{2025ApJ...983...76H}). This leads to significant variations in the object's apparent brightness without any intrinsic changes of the physical
processes in the binary system.
For example the recent Cousins I ($800\pm150$~nm) lightcurve of
T~CrB, which can be obtained from the AAVSO database \citep{aavso}, has a near-sinusoidal shape with a period of half the orbital period of T~CrB and an amplitude (in flux density) of ca. 30\% of the maximum flux density.
We must assume that also the emission at mm wavelengths is modulated by the
binary's rotation, not necessarily with the same relative amplitude as
in the optical, but with the same period. 
And if the amplitude of the modulation is wavelength-dependent, this results in a modulation of the spectral shape.

The second main reason for expecting the mm emission of T~CrB to have a variable spectrum
is that the accretion rate onto the WD is variable leading to moderate
high-states which have been studied, as mentioned in the introduction, at
multiple wavelengths, also by \citet[][hereafter L2019]{2019ApJ...884....8L} in the frequency window just below that of ALMA
at 5~GHz to 35~GHz.

   \begin{figure}
   \resizebox{\hsize}{!}{\includegraphics{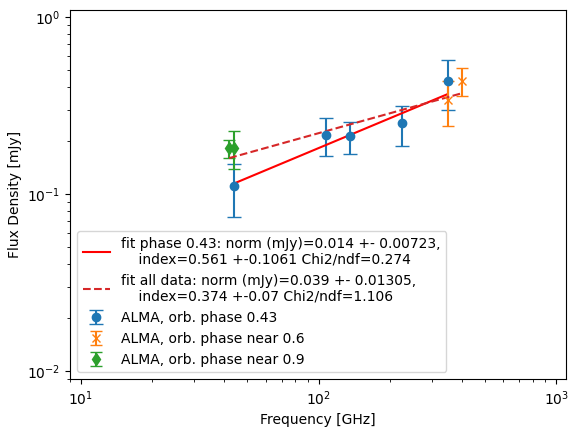}}
   \caption{Our 2024 ALMA T~CrB flux density measurements from Table~\ref{table:fluxes} plotted vs. centroid frequency and coloured by orbital phase (see legend). The error bars include the systematic errors.
    The solid line shows a power law fit to the points for orbital phase 0.43
    while the dashed line shows a power law fit to all points.
    }
              \label{fig:alma-spec}%
    \end{figure}

Our pre-burst observations of T~CrB span nearly half a rotation of the system (46\% from phase 0.4328 to 0.8889) but we are far from having a complete
coverage of all bands at all orbital phases.
We only have one nearly complete set of
observations at phase 0.43 with bands 1, 3, 4, 6, and 7. 
Figure \ref{fig:alma-spec} shows a power-law fit (solid line) to these
five flux density measurements we have at near-equal phase (see 
also Table~\ref{table:fluxes}).
Super-imposed, but not included in the fit, are the remaining measurements
from Table~\ref{table:fluxes} which were made at other orbital phases.
If we include all nine measurements in the fit, we obtain the dashed line.
The best fit parameters are given in Table~\ref{tab:specfit}. All fits were performed under Python~3 using the "curve\_fit" function from the scipy.optimize package (see https://scipy.org).
\begin{table}
    \centering
    \caption{Results of fits of a power law ($F(\nu) = F_0 \nu^{\alpha}$, $\nu$~(GHz))
    to different subsets of our data points from Table \ref{table:fluxes}.}
    \label{tab:specfit}
\begin{tabular}{l l c c}
\hline\hline
orbital phase & Norm $F_0$(mJy) & spectral index $\alpha$ & $\chi^2/\mathrm{ndf}$\\
\hline
only 0.43 &  0.014$\pm$0.0072 &  0.56$\pm$0.11 & 0.27\\
all phases &  0.039$\pm$0.013 &  0.374$\pm$0.070 & 1.11\\
\hline
\end{tabular}

\end{table}

While the reduced $\chi^2$ of 1.1 for the power law fit to all data points is
perfectly acceptable, and the spectral index is formally better constrained
when all data are used, we note that due to the well motivated expectation
of spectral variability, only our spectrum for phase 0.43 can be regarded 
as well defined because our dataset lacks spectral coverage at other phases.
However, due to the faintness of the target, both spectra are still marginally
consistent.

   \begin{figure}
   \resizebox{\hsize}{!}{\includegraphics{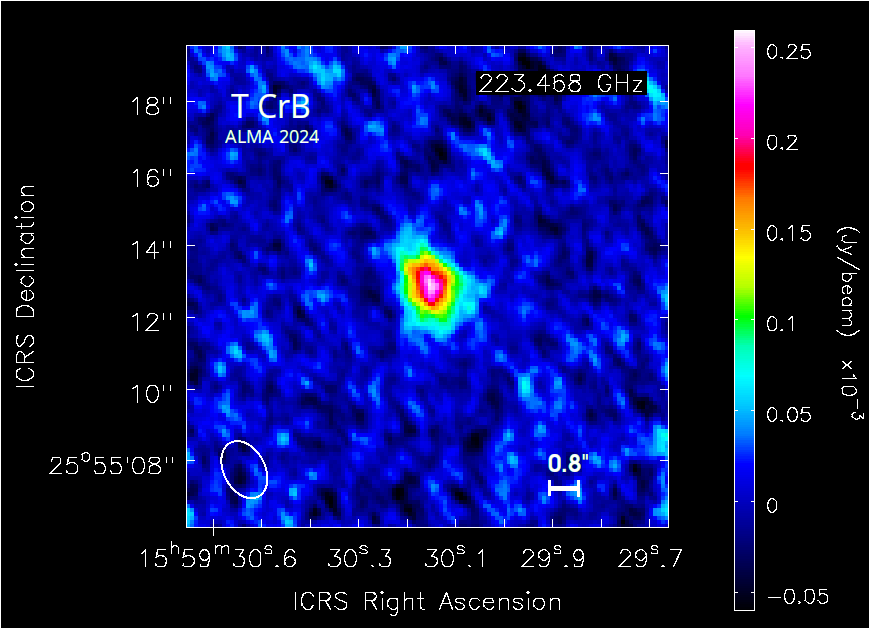}}
   \caption{Wide-band image of T~CrB obtained from a joint MTMFS deconvolution
       of all our ALMA 12~m array data on the target (bands 1 - 8) from 2024 as 
       listed in Table~\ref{table:obs}. The size of the synthesized beam 
       (angular resolution) is given by the ellipse plotted in the lower left
       corner. The parameters (major axis, minor axis, position angle) are
       1.62~arcsec, 1.16~arcsec, 19.8$^\circ$. The image resembles that of the point spread function. The object is detected with 15~$\sigma$ significance but there are
       no resolved spatial features. ALMA was tracking the object at all
       times taking into account its proper motion. The spatial coordinates
       shown are those for 10 August 2024, the start of the observations.}
              \label{fig:image}%
    \end{figure}

To help constrain the spectrum further and at the same time look for
potential signs of extended structure (our observations in bands 6 - 8
have angular resolutions better than 0.5~arcsec), we combine all our ALMA 12~m 
array data and image it using CASA tclean with deconvolver MTMFS 
with two Taylor terms (nterms=2) 
such that in addition to a high-resolution image, we also obtain
an estimate of the best fit power law spectral index of the object.
The resulting image is shown in Fig. \ref{fig:image}. It is in good
agreement with the point spread function of the dataset, and thus there
is no indication of any resolved spatial features. 

The image deconvolution is unproblematic and converges quickly. 
The achieved noise RMS is 0.018~mJy/beam, the peak flux density is 0.258~mJy/beam,
and the centroid frequency of the observation is 223.468~GHz.
The spatial century coordinates of the image are chosen to be those for the
start of the observation on 10 August 2024. ALMA was, however,
tracking the object at all times taking into account its proper motion.
So the image center coincides with the nominal position of T~CrB.
Fitting a Gaussian to the emission results in a peak position 
which is inside the central pixel, confirming the correctness
of the ephemeris of T~CrB within the angular resolution.
Also in all individual per-band images, the peak position was
less than a beam radius away from the image center.

The primary-beam-corrected "alpha" and "alpha.error" images (spectral index $\alpha$ 
and corresponding error estimate)
produced by the MTMFS deconvolution contain in their central pixels
(where also the peak flux density is measured) the value 
$\alpha=0.57\pm0.23$
which is in very good agreement with our measurement in Table~\ref{tab:specfit} albeit with a larger uncertainty.

   \begin{figure}
    \resizebox{\hsize}{!}{\includegraphics{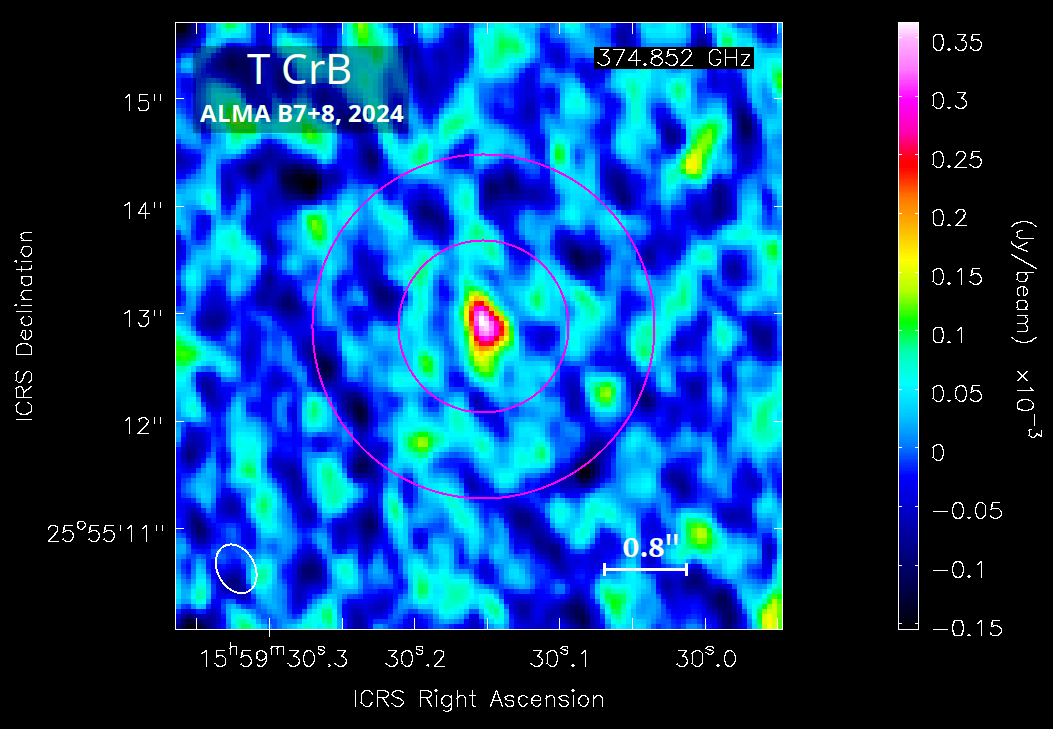}}
   \caption{Continuum image obtained from combining only our band 7 and 8 observations of T~CrB where we achieve the best angular resolution. The two magenta circles represent the boundaries of an annulus around the position of T~CrB with inner radius 0.8~arcsec (twice the angular resolution of the observation) and outer radius 1.6~arcsec, inside which we derive an upper limit on diffuse emission in the circum-binary medium.}
              \label{fig:trc-b78-cont}
    \end{figure}

   \begin{figure}
    \resizebox{\hsize}{!}{\includegraphics{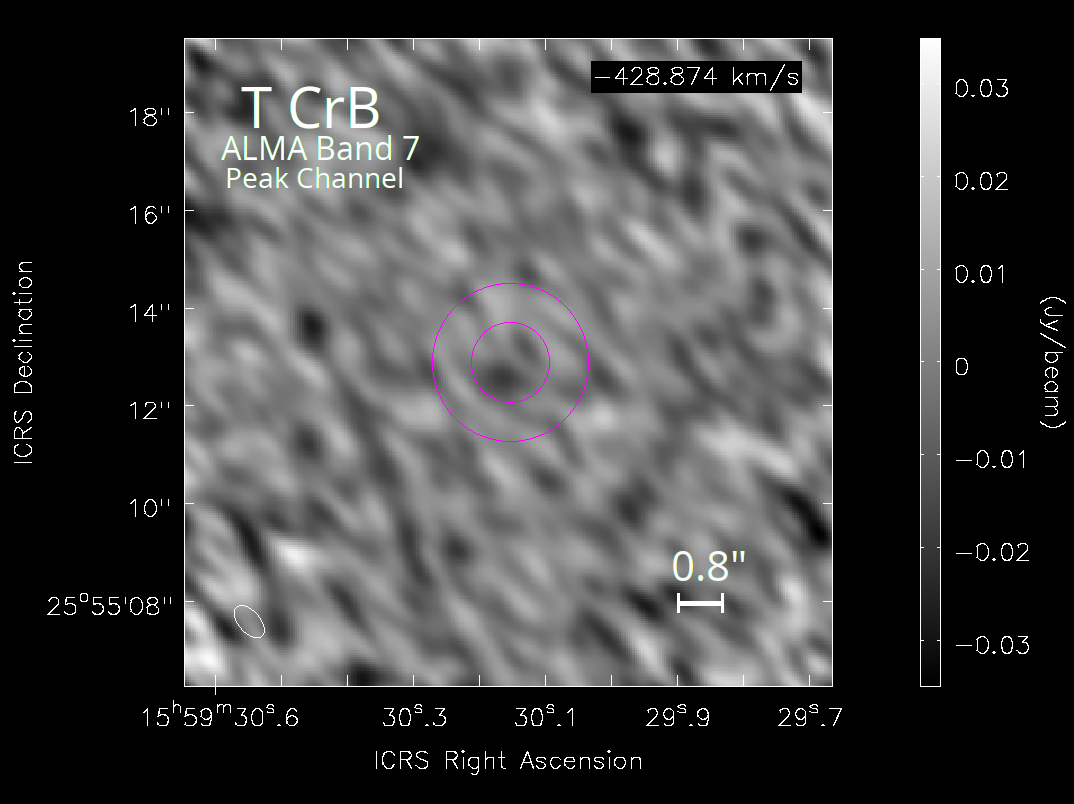}}
   \caption{Image from the peak channel of spectral window 4 in our complete ALMA band 7 observations of the region around T~CrB in 2024 (see Tab.~\ref{table:speclimits}), where we achieve the best angular resolution and sensitivity for an individual line. This spectral window is centered (in LSRK) on the rest frequency of HCN v=0 J=4-3 line. The radio velocity shown in the upper right is
   relative to this line. The image is consistent with pure noise.
   The signal, for which this image shows the peak channel, is extracted from the annulus given by the two magenta circles around
   the image center, which are the same as in Fig. \ref{fig:trc-b78-cont}. The beam major and minor axes are 0.48~arcsec and 0.30~arcsec. We show this image in gray-scale to draw attention the fact that this is a narrow-band image of a single spectral channel.}
              \label{fig:trc-b7-peak}
    \end{figure}

The gas content of the circum-binary region of T~CrB is best constrained by our band 7 and 8 observations because here we achieve the best angular resolution. In the combined image from all our band~7 and 8 data, the beam FWHM major and minor axes are 0.48~arcsec and 0.35~arcsec, 0.42~arcsec on average (Fig.~\ref{fig:trc-b78-cont}). In an annulus of 0.8~arcsec (= twice the angular resolution to avoid any residual sidelobes of the point spread function) inner radius and 1.6~arcsec outer radius centered on T~CrB (equivalent to 720~AU - 1440~AU at the distance of T~CrB based on Gaia DR3 data \citep{2021AJ....161..147B}), we obtain a mean surface brightness of 4.2~$\mu$Jy/beam and an RMS of 43.7~$\mu$Jy/beam. This means there is no significant continuum emission between ca. 344~GHz and 407~GHz down to a surface brightness of 0.09~mJy/beam with a confidence level of 2~$\sigma$. With the given beam size, this corresponds to a brightness temperature upper limit in the annulus of $T_b < 4.7$~K.

The line emission in this spectral and spatial region is best constrained in band 7 (7 and 7b combined), spectral window 4 (velocity resolution 0.82~km/s, see also Tab.~\ref{table:speclimits}). The most prominent potential line in this window is the tracer HCN v=0 J=4-3 with a rest frequency near 354.5055~GHz. Neither this nor any other line is detected. Even if the frequency of the line was shifted differently at different orbital periods (since we combine 7 and 7b with corresponding orb. phases 0.43 and 0.59), any upper limit on the HCN line emission must be $\leq$ that 
obtained from the channel with the brightest signal.
We thus obtain a $2 \sigma$ upper limit on the line surface brightness (assuming a 10~km/s line width) for any potential line in spectral window 4 from the peak channel of that spectral window with a value of 3.1~mJy/beam. 
There is no indication of any spatial structure. The image of the peak and all other channels look like noise (Fig. \ref{fig:trc-b7-peak}).

\section{Spectral re-analysis of the 2016/17 VLA data}
\label{sec-vlaspec}
Since our ALMA data is the first measurement of T~CrB at frequencies between
42~GHz and 407~GHz, we cannot make statements about long-term variability
in this regime. We can, however, assume a smooth spectral transition to
the frequency window just below ours and compare to the measurements made
by L2019 in 2016/17 with the VLA at 5~GHz to 35~GHz.

   \begin{figure}
    \resizebox{\hsize}{!}{\includegraphics{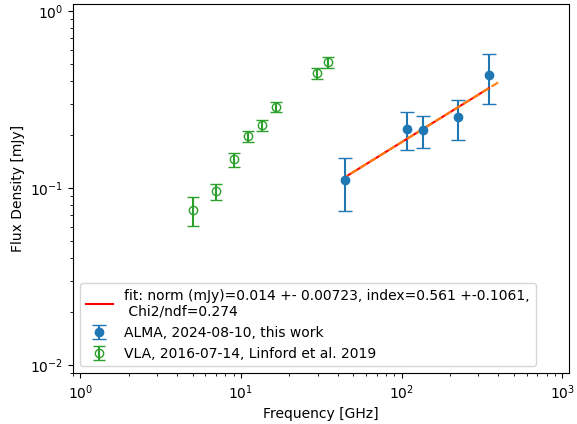}}
   \caption{Comparison of our ALMA spectrum of T~CrB for orbital phase 0.43
     from 10 Aug 2024 (blue points) with a representative spectrum of the same object
     observed on 14 July 2016 with the VLA by \cite{2019ApJ...884....8L} (green circles). The fit parameters given in the caption are those of a power law fit
     to our points as also shown in Fig. \ref{fig:alma-spec}.}
              \label{fig:joint-spec}%
    \end{figure}
    
Figure \ref{fig:joint-spec} shows our spectrum for phase 0.43 (August 2024) together with a representative
spectrum from L2019 for the date 14~July~2016 where they have a complete
set of data points over their full frequency range.
In 2016/17, T~CrB was in a high state, and (as one can see from Fig. \ref{fig:joint-spec} by extrapolating our spectrum down in frequency by a few GHz) approximately a factor 5 brighter around 35~GHz
than it was in the second half of 2024 when ALMA observed it.
The decrease in brightness at IR or visible wavelengths between 2016/17 and 2024 was much more modest
(ca. 0.5 mag corresponding to a factor 1.58) according to the AAVSO database \citep{aavso}. 

Furthermore, the T~CrB spectrum from 2016/17 with an average index of 1.0$\pm$0.03 (from 5~GHz to 35~GHz, averaging over values given by L2019 for
all their individual nights)
was significantly harder than our spectrum with index 0.56$\pm$0.11 (also shown in  Fig. \ref{fig:joint-spec}).

A closer look at the seven individual spectra presented by L2019 for their
different visits to T~CrB in 2016/17 reveals that the spectra may have some curvature. 
L2019 do not mention the goodness of their powerlaw fits but  if we re-fit their four points above 13.5~GHz in each of their spectra
with a separate power-law, we find that in all cases except two, the fit is good and results in
an index between 0.82 and 1.0 . The exceptions are the nights of 2016-08-25 and 2016-09-22, where the obtained index is much lower (0.29 and 0.51 respectively) with worse
goodness of fit. Inspecting the "Observations" section of L2019, one finds on pages 2 and 3 of the paper that
(a) the two points from August and September 2016 were the only ones where the VLA atmospheric	delay model "caused issues with source location and smearing along the direction of elevation" (although L2019 argued that the atmospheric delay model issue did not affect their measurements).
(b) The above effect was larger for higher frequencies.
(c) The 29.5~GHz and 35~GHz points for these two months were "found to be decorrelated". 
The checksource is used to derive a correction to the flux density. The error bars of the 
measurements were "increased accordingly". We interpret this as a justification to exclude the two
observing nights from the sample. They obviously are affected by systematic
errors which make them not useful for our investigation of the variability of
the spectral index.

   \begin{figure}
    \resizebox{\hsize}{!}{\includegraphics{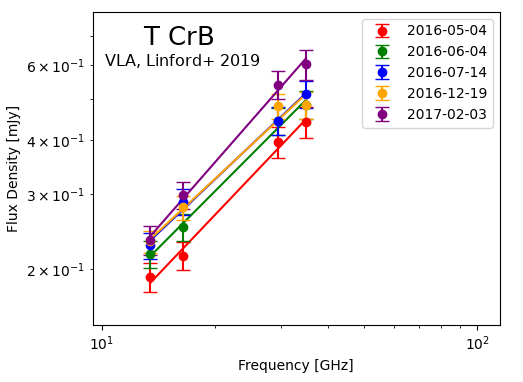}}
   \caption{The VLA data points from \cite{2019ApJ...884....8L} used by us to derive the 
      spectral indices in Table~\ref{tab:vlaspectra}. The lines show our
      power law fits.}
              \label{fig:vlaspectra}%
    \end{figure}

\begin{table}
    \centering
   \caption{Results from our powerlaw fits to the upper four points (13.5~GHz to 35~GHz)
   of the per-visit spectra of T~CrB by \cite{2019ApJ...884....8L} observed with the VLA in 2016/17.}
    \label{tab:vlaspectra}
\begin{tabular}{c c c c}
\hline\hline
MJD & orbital phase & spectral index $\alpha$ & $\chi^2/\mathrm{ndf}$\\
\hline
57512.39 & 0.891 & 0.921$\pm$ 0.056 & 0.31\\
57543.33 & 0.028 & 0.880$\pm$ 0.040 & 0.19\\
57583.18 & 0.203 & 0.829$\pm$ 0.045 & 0.25\\
57741.76 & 0.900 & 0.817$\pm$ 0.077 & 0.86\\
57787.69 & 0.102 & 1.001$\pm$ 0.043 & 0.19\\
\hline
\end{tabular}
 
\end{table}

Table \ref{tab:vlaspectra} lists our fit results for the L2019 data
between 13.5~GHz and 35~GHz on the five nights with good, comparable data.
The average spectral index for these nights is 0.90$\pm$0.02 (13.5~GHz - 35~GHz)
compared to the average spectral index 0.96$\pm0.03$ measured by L2019 over their full 
spectral range 5~GHz to 35~GHz for the same nights.

Comparing the different spectral indices obtained from the VLA data from 2016/17
in the spectral range just below ours with the spectral index obtained from our observations
in 2024, we can say that T~CrB very likely had a softer
spectrum around 40~GHz in 2024 than it had at any time during the VLA observations.
However, the difference between the softest VLA index and ours is only 0.26$\pm$0.13,
a 2 sigma effect. Deeper observations covering the entire orbital period
would be needed to draw firmer conclusions. 
Compared to the average VLA index we measure between 13.5 and 35~GHz (see above), 
our ALMA index is softer by 0.34$\pm$0.11 .

\section{Discussion}
\label{sec-modelfit}

\subsection{Model}

In order to interpret the VLA and ALMA spectra, we work with the simplified model inherited from \cite{1975A&A....39....1P} 
and  \cite{1975MNRAS.170...41W}: 
an isotropic, isothermal, fully ionized, and free-free emitting stellar wind. The radiation from the WD and its accretion disk are responsible for the anomalously high wind ionization and temperature compared to an isolated red giant star. Yet, we assume the wind temperature $T_0$ to be uniform and discard geometrical effects such as the ones considered by \cite{1984ApJ...286..263T}. 

The wind number density profile $n(r)$ is set to a power-law:
\begin{equation}
\label{eq:n}
    n(r)=n_0\left(\frac{r}{R}\right)^{-\beta}
\end{equation}
with $\beta>0$ the density profile index, a dimensionless constant, $R$ the stellar radius and $n_0$ the number density scale. In these conditions, there is a critical frequency $\nu_t$, called the turnover frequency, above which the wind is fully optically thin. The behavior of the flux density $S_{\nu}$ is twofold, depending on whether $\nu\gg\nu_t$ (optically thin regime) or $\nu<\nu_t$ (optically thin and thick hybrid regime).

In the frequency range we span (5~GHz to 400~GHz) and for realistic wind temperatures ($T_0>10^4$~K such that the fully ionized hypothesis holds), we can make the Rayleigh-Jeans approximation ($h\nu\ll k_BT_0$) and use a simplified form of the free-free opacity in order to derive analytical expressions. In these conditions, the spectral index $\alpha$ of the free-free emission of the wind is $\sim -0.1$ in the optically thin regime. In the hybrid regime, below the turnover frequency, the spectral index $\alpha$ is uniquely set by the exponent $\beta$ in equation\,\eqref{eq:n} through:
\begin{equation}
\label{eq:p_indx}
    \alpha\sim2-\frac{4.2}{2\beta-1}
\end{equation}
such that a wind at uniform speed (which means $\beta=2$ from mass conservation) produces a spectral index of $\alpha\sim0.6$. However, since the wind steadily accelerates from the dust condensation radius (the distance from the star where the temperature is low enough for dust grains to form and couple to the underlying continuum, \citealt{2018A&ARv..26....1H}), we expect $\beta>2$, at least up to a distance where the wind has reached its terminal speed. This means that accounting for wind acceleration yields a steeper spectrum, with $\alpha>0.6$.

In radio/mm, an accurate estimate of the turnover frequency can be obtained from the implicit equation:
\begin{equation}
\label{eq:nut}
    \nu_t\sim 7.7\cdot10^{-2}I(\beta)n_0R^{1/2}T_0^{-3/4}f_g^{1/2}(\nu_t,T_0) \,\mathrm{Hz}
\end{equation}
where:
\begin{equation}
    I(\beta)=\int_0^{+\infty}\frac{dx}{(1+x^2)^{\beta}}
\end{equation}
and $f_g$ is the Gaunt factor, a slowly varying function of $\nu$ given by \cite{2014MNRAS.444..420V}:
\begin{equation}
    f_g(\nu,T)\sim10.6+1.9\log T-1.26\log \nu \quad \text{(in CGS units)}
\end{equation}
Therefore, once the exponent $\beta$ has been obtained from the spectral index in the hybrid regime, a measure of the turnover frequency provides a relation between $n_0$, $R$ and $T_0$.

Finally, the norm $K$ of the spectral fit in the hybrid regime is given by:
\begin{equation}
\label{eq:K}
    K\sim \pi \left[\frac{I(\beta)}{3}\right]^{\frac{2}{2\beta-1}}\frac{\Sigma(\beta)}{D^2\nu_0^{\alpha}}B_{\nu_0}(T_0)n_0^{\frac{4}{2\beta-1}}R^{2+\frac{2}{2\beta-1}}\chi_{\nu_0}(T_0)^{\frac{2}{2\beta-1}}
\end{equation}
where $D$ is the distance to the source, $\nu_0$ is a frequency representative of the frequency range. The exact choice of $\nu_0$ has only a minor impact on the results, and so we set it to $30$~GHz,  essentially the geometric mean of the VLA-ALMA frequency range. $B_{\nu}(T)$ is the Planck function, $\Sigma(\beta)$ is a series given by:
\begin{equation}
    \Sigma(\beta)=1+2\sum_{n=1}^{+\infty}\frac{(-1)^{n+1}}{n!}\frac{3^n}{(2\beta-1)n-2}
\end{equation}
and $\chi_{\nu}(T)$ is given by
\begin{align}
    \chi_{\nu}(T)&=\frac{\kappa_{ff}}{n^2}\sim 3.692\cdot 10^8 \left(1-e^{-h\nu/k_BT}\right) T^{-1/2} \nu ^{-3} f_g(\nu,T) \\
    &\sim 1.772\cdot 10^{-10}T^{-3/2}\nu^{-2}f_g(\nu,T) \quad \text{if $h\nu\ll k_BT$}
\end{align}
(in CGS units) with $\kappa_{ff}$ the free-free opacity. Therefore, if we know the distance $D$ and the exponent $\beta$, the norm $K$ provides a second relation between $n_0$, $R$ and $T_0$.

\begin{figure*}
\centering
 \resizebox{\hsize}{!}{\includegraphics{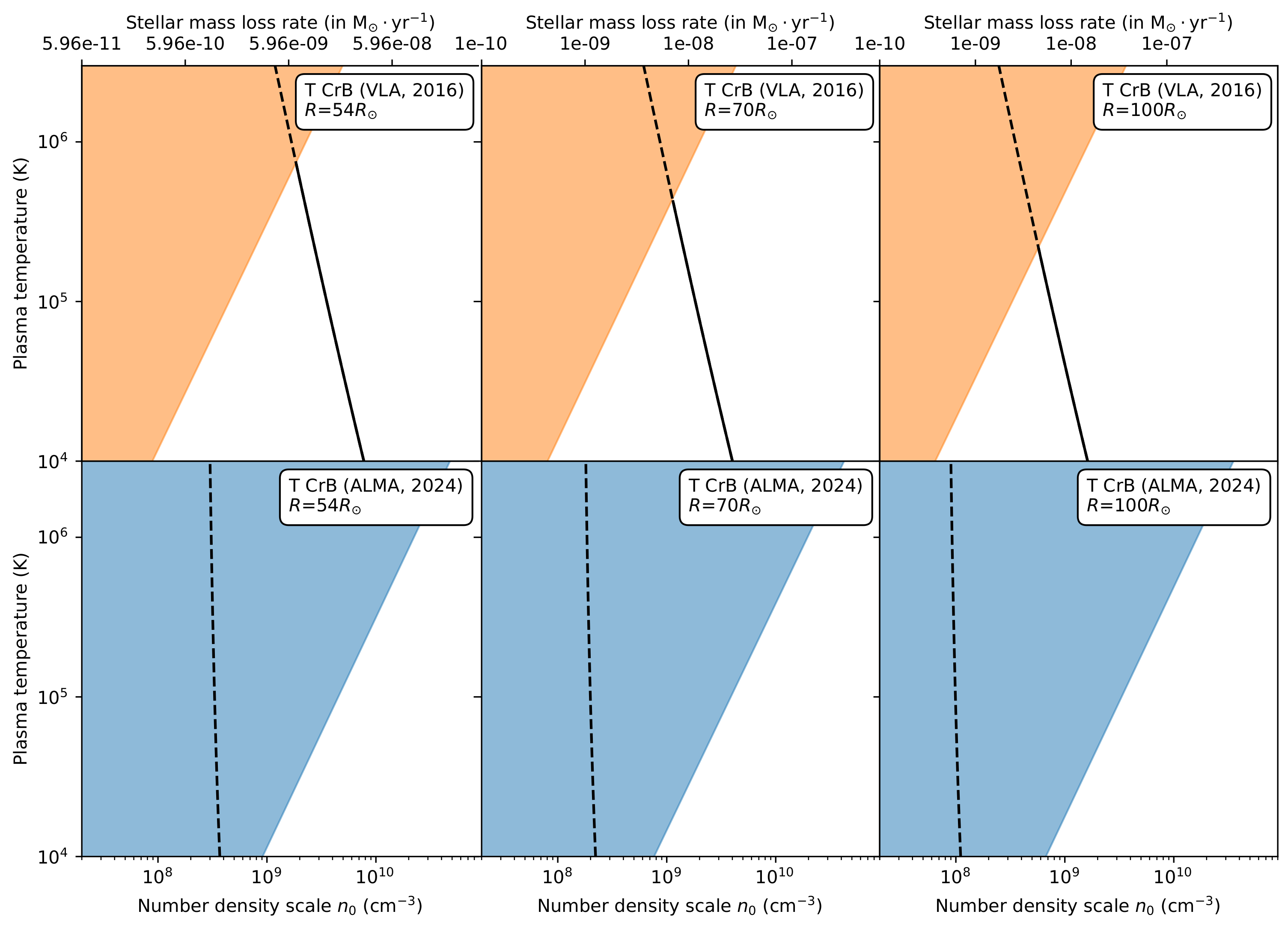}}
\caption{Possible values of red giant wind particle number density scale and plasma temperature $(n_0,T_0)$ (solid lines) for the 2016 VLA (top panels) and 2024 ALMA spectra (bottom) of T~CrB, for an assumed stellar radius of $R=54~R_{\odot}$ (left), $R=70~R_{\odot}$ (middle) and $R=100~R_{\odot}$ (right). The regions allowed by the lower limit on the turnover frequency are in white. Inside the shaded regions, the lines of $(n_0,T_0)$ solutions are shown as dashed in order to indicate that they do not respect the constraint set by the lower limit on the turnover frequency and are thus not physical. The top axis indicates stellar mass-loss rates corresponding to the number density $n_0$. A fiducial frequency $\nu_0=30$~GHz, whose choice little alters the result compared to other systematics, was used to compute $T_0(n_0)$ from equation\,\eqref{eq:K}.
}
\label{fig:n0T0}%
\end{figure*}

\subsection{Data interpretation}

The spectral indices we find for the ALMA and VLA data (Tables \ref{tab:specfit} and \ref{tab:vlaspectra}) are all incompatible with the fully optically thin regime. While there is a suggestion of some curvature in the VLA spectra, we do not see a significant break which would suggest that the turnover frequency is well inside our frequency range. Hereafter, we set a lower limit to the turnover frequency to the last data point of each dataset:
\begin{align}
    &\nu_{\text{t,min},\text{2016/17}}=35~\text{GHz}\\
    &\nu_{\text{t,min},\text{2024}}=350~\text{GHz}
\end{align}
and we interpret this data in the hybrid regime ($\nu<\nu_t$). With equation\,\eqref{eq:p_indx} 
and the spectral indices we fitted, we deduce:
\begin{align}
\label{eq:beta1}
    &\beta_{\text{2016/17}} = 2.5 \pm 0.09\\
\label{eq:beta2}
    &\beta_{\text{2024}} = 2.0 \pm 0.11 
\end{align}
To derive the uncertainty estimates given in \eqref{eq:beta1} and \eqref{eq:beta2}, we solved equation \eqref{eq:p_indx} for $\beta$ and then propagated the uncertainty values of spectral index $\alpha$ from Tables~\ref{tab:specfit} and \ref{tab:vlaspectra} respectively into an uncertainty of the corresponding $\beta$. The derived uncertainties of the $\beta$ values do not include systematics arising from the simplifying assumptions of the model (e.g. assuming a power-law for the wind density profile). These systematics would probably dominate but are not random and would shift both derived values of $\beta$ in the same direction. So while the absolute values of $\beta$ are 
quite uncertain, we can argue that they are different by more than 3~$\sigma$,
meaning that the underlying wind dynamics were different in the two time periods, 
with a more accelerating wind in 2016/17 and
a more uniform wind speed profile from its basis during our 2024 observations.

Since this data spans the hybrid regime, we only have the lower limits on the turnover frequency $\nu_t$, which translate into lower limits for $n_0$ and $R$, and an upper limits for $T_0$. Then, equation\,\eqref{eq:nut} delimits a range of possible triplets $(n_0,R,T_0)$.

Now, we use equation\,\eqref{eq:K} to obtain a relation between $n_0$, $R$ and $T_0$, knowing $\beta$ and using the distance $D\sim$895~pc \citep{2021AJ....161..147B}. 
From the spectral type of the donor star which is approximately M2 III to M4.5 III 
\citep{1993A&A...270..165Y, 1999A&AS..137..473M, 2016NewA...47....7M, 2023A&A...680L..18Z} 
and from arguments related to mass transfer through Roche lobe overflow 
\citep{2023A&A...680L..18Z,2024AN....34540036Z,2025A&A...694A..85P},
we expect a stellar radius ranging between 70 and 100~$R_{\odot}$ according to the evolutionary tracks by  \cite{1981Ap&SS..80..353S}. The latest analysis by \cite{2025ApJ...983...76H} suggest, however, that a
radius of 63.4~$R_\odot$ is more correct (assuming the Gaia DR3 distance) or an even lower value of 54~$R_\odot$ if the distance is made a fit parameter (see introduction).  
We plot the relation $T_0(n_0)$ from $K$ for the two extreme potential values of $R$ and for 70~$R_\odot$, and we retain only the values which match the constraint from the corresponding turnover frequency's lower limit. 

Results are shown in Figure\,\ref{fig:n0T0}, where allowed values lie in the white regions. In the top $x$-axis, we provide an estimate of the stellar mass-loss rate assuming a wind launched at $10$~km/s from the stellar surface (and neglecting the inherent inconsistencies due to $\beta>2$). We see that for the 2016/17 VLA data, this model gives a range of possible values. The stellar mass loss rate we measure is in the range 1 to 5$\times 10^{-8}$M$_{\odot}/$yr, in agreement with the expected values from the accretion rate during the high-state \citep{2004A&A...415..609S}
and from the recurrence period of the nova \citep{2023JHA....54..436S}. 
However, the 2024 ALMA data cannot be interpreted with this model. Indeed, the flux is too low and the turnover frequency too high to yield a range of possible values for $n_0$ and $T_0$, for any realistic stellar radius $R$. The line of possible $(T_0,n_0)$ values is fully inside the excluded region in the three lower plots in Fig.~\ref{fig:n0T0}. We think that this is due to the low activity of the system in 2024 compared to 2016/17, which would mean that the ionizing flux from the accretion disk was much lower and only a small fraction of the stellar wind, centered on the white dwarf, was sufficiently ionized to radiate through free-free emission. 
    
\section{Conclusions}
\label{sec-conc}
We analyzed our ALMA 2024 T~CrB data and for comparison VLA 2016/17 data by \cite{2019ApJ...884....8L} in order to determine the properties (density, temperature and ionization) of the stellar wind in which the upcoming nova will take place. 
We find that the spectral index above 35~GHz in 2024 is significantly lower than the index measured just below 35~GHz in 2016/17.
But in both cases the index is well above the optical thin regime ($-0.1$). So, both in 2016/17 and in 2024, the turnover frequency $\nu _t$, above which the wind becomes fully optically thin, was well above 35~GHz, and in 2024 it was even above 350~GHz.
It is unlikely that the spectrum at ALMA energies was the same in 2016/17 and in 2024, because it would not fit the expected shape of the turnover, which should happen within a decade of frequency.
What we see at ALMA energies, however, is consistent with a straight power-law from 40~GHz to 400~GHz.
We conclude that the overall state of the object and its wind must have been very different in 2016/17 and 2024, and the spectra of the two states are not just scaled versions of each other with different intensity scales.

We showed that the VLA 2016/17 data can be well explained by radio free-free emission from an isotropic, isothermal and fully ionized wind associated with a stellar mass loss rate of $\sim 10^{-8}$M$_\odot$/yr. However, this model is unable to account for the ALMA 2024 low flux and high turnover frequency, which suggests that a more complex model (beyond the scope of this paper) needs to be employed for the 2024 state involving an only partially ionized wind. This is in agreement with the reported end of the super-active phase in early 2023 \citep{2023RNAAS...7..145M}. 

Our aggregate bandwidth images of the object in all ALMA bands are consistent with that of a point source.
The most stringent limit on the presence of matter in the space near the object is achieved in bands 7 and 8 with a continuum sensitivity of 0.09~mJy/beam limiting the brightness temperature near 374~GHz to $T_b<4.7$~K in a 720~AU to 1440~AU annulus (assuming the Gaia DR3 distance) around T~CrB. The emission observed from the central point source is constrained to an angular radius of 0.4~arcsec (corresponding to 360~AU assuming the same distance). No line emission is detected in any band.

In future ALMA observations of the quiescent state of T~CrB and similar objects, the turnover frequency could be further constrained by extending the concurrent frequency coverage (within a few days) into band 9 and doubling the flux sensitivity.

\begin{acknowledgements}
      This paper makes use of the following ALMA data: ADS/JAO.ALMA\#2023.A.00038.T, ADS/JAO.ALMA\#2023.A.00041.S. ALMA is a partnership of ESO (representing its member states), NSF (USA) and NINS (Japan), together with NRC (Canada), NSTC and ASIAA (Taiwan), and KASI (Republic of Korea), in cooperation with the Republic of Chile. The Joint ALMA Observatory is operated by ESO, AUI/NRAO and NAOJ. GS acknowledges support by the Spanish MINECO grant PID2023-148661NB-I00, by the E.U. FEDER funds, and by the AGAUR/Generalitat de Catalunya grant SGR-386/2021. GS acknowledges support from the Max-Planck-Institut f\"ur extraterrestrische Physik for her visit in summer 2024, when fruitful discussions about T~CrB and ALMA with the coauthors gave rise to the present work. IEM acknowledges support from the grant ANID FONDECYT 11240206 and funding via the BASAL Centro de Excelencia en Astrofisica y Tecnologias Afines (CATA) grant PFB06/2007. We acknowledge useful discussions with Steven N. Shore (Universit\'{a} de Pisa) and the anonymous referee.
\end{acknowledgements}


\begin{thebibliography}{51}
\expandafter\ifx\csname natexlab\endcsname\relax\def\natexlab#1{#1}\fi

\bibitem[{{Acciari} {et~al.}(2022){Acciari}, {Ansoldi}, {Antonelli}, {Arbet Engels}, {Artero}, {Asano}, {Baack}, {Babi{\'c}}, {Baquero}, {Barres de Almeida}, {Barrio}, {Batkovi{\'c}}, {Becerra Gonz{\'a}lez}, {Bednarek}, {Bellizzi}, {Bernardini}, {Bernardos}, {Berti}, {Besenrieder}, {Bhattacharyya}, {Bigongiari}, {Biland}, {Blanch}, {B{\"o}kenkamp}, {Bonnoli}, {Bo{\v{s}}njak}, {Busetto}, {Carosi}, {Ceribella}, {Cerruti}, {Chai}, {Chilingarian}, {Cikota}, {Colak}, {Colombo}, {Contreras}, {Cortina}, {Covino}, {D'Amico}, {D'Elia}, {Da Vela}, {Dazzi}, {De Angelis}, {De Lotto}, {Del Popolo}, {Delfino}, {Delgado}, {Delgado Mendez}, {Depaoli}, {Di Pierro}, {Di Venere}, {Do Souto Espi{\~n}eira}, {Prester}, {Donini}, {Dorner}, {Doro}, {Elsaesser}, {Fallah Ramazani}, {Fari{\~n}a Alonso}, {Fattorini}, {Fonseca}, {Font}, {Fruck}, {Fukami}, {Fukazawa}, {Garc{\'\i}a L{\'o}pez}, {Garczarczyk}, {Gasparyan}, {Gaug}, {Giglietto}, {Giordano}, {Gliwny}, {Godinovi{\'c}}, {Green}, {Green}, {Hadasch}, {Hahn}, {Hassan}, {Heckmann},
  {Herrera}, {Hoang}, {Hrupec}, {H{\"u}tten}, {Inada}, {Ishio}, {Iwamura}, {Jim{\'e}nez Mart{\'\i}nez}, {Jormanainen}, {Jouvin}, {Kerszberg}, {Kobayashi}, {Kubo}, {Kushida}, {Lamastra}, {Lelas}, {Leone}, {Lindfors}, {Linhoff}, {Lombardi}, {Longo}, {L{\'o}pez-Coto}, {L{\'o}pez-Moya}, {L{\'o}pez-Oramas}, {Loporchio}, {Machado de Oliveira Fraga}, {Maggio}, {Majumdar}, {Makariev}, {Mallamaci}, {Maneva}, {Manganaro}, {Mannheim}, {Maraschi}, {Mariotti}, {Mart{\'\i}nez}, {Mas Aguilar}, {Mazin}, {Menchiari}, {Mender}, {Mi{\'c}anovi{\'c}}, {Miceli}, {Miener}, {Miranda}, {Mirzoyan}, {Molina}, {Moralejo}, {Morcuende}, {Moreno}, {Moretti}, {Nakamori}, {Nava}, {Neustroev}, {Nievas Rosillo}, {Nigro}, {Nilsson}, {Nishijima}, {Noda}, {Nozaki}, {Ohtani}, {Oka}, {Otero-Santos}, {Paiano}, {Palatiello}, {Paneque}, {Paoletti}, {Paredes}, {Pavleti{\'c}}, {Pe{\~n}il}, {Persic}, {Pihet}, {Prada Moroni}, {Prandini}, {Priyadarshi}, {Puljak}, {Rhode}, {Rib{\'o}}, {Rico}, {Righi}, {Rugliancich}, {Sahakyan}, {Saito}, {Sakurai},
  {Satalecka}, {Saturni}, {Schleicher}, {Schmidt}, {Schweizer}, {Sitarek}, {{\v{S}}nidari{\'c}}, {Sobczynska}, {Spolon}, {Stamerra}, {Stri{\v{s}}kovi{\'c}}, {Strom}, {Strzys}, {Suda}, {Suri{\'c}}, {Takahashi}, {Takeishi}, {Tavecchio}, {Temnikov}, {Terzi{\'c}}, {Teshima}, {Tosti}, {Truzzi}, {Tutone}, {Ubach}, {van Scherpenberg}, {Vanzo}, {Vazquez Acosta}, {Ventura}, {Verguilov}, {Vigorito}, {Vitale}, {Vovk}, {Will}, {Wunderlich}, {Yamamoto}, {Zari{\'c}}, \& {Ambrosino}}]{2022NatAs...6..689A}
{Acciari}, V.~A., {Ansoldi}, S., {Antonelli}, L.~A., {et~al.} 2022, Nature Astronomy, 6, 689

\bibitem[{{Azzollini} {et~al.}(2022){Azzollini}, {Shore}, \& {Kuin}}]{2022RNAAS...6...92A}
{Azzollini}, A., {Shore}, S.~N., \& {Kuin}, P.~N. 2022, Research Notes of the American Astronomical Society, 6, 92

\bibitem[{{Azzollini} {et~al.}(2023){Azzollini}, {Shore}, {Kuin}, \& {Page}}]{2023A&A...674A.139A}
{Azzollini}, A., {Shore}, S.~N., {Kuin}, P.~N., \& {Page}, K.~L. 2023, A\&A, 674, A139

\bibitem[{{Bailer-Jones} {et~al.}(2021){Bailer-Jones}, {Rybizki}, {Fouesneau}, {Demleitner}, \& {Andrae}}]{2021AJ....161..147B}
{Bailer-Jones}, C.~A.~L., {Rybizki}, J., {Fouesneau}, M., {Demleitner}, M., \& {Andrae}, R. 2021, \aj, 161, 147

\bibitem[{{CASA Team} {et~al.}(2022){CASA Team}, {Bean}, {Bhatnagar}, {Castro}, {Donovan Meyer}, {Emonts}, {Garcia}, {Garwood}, {Golap}, {Gonzalez Villalba}, {Harris}, {Hayashi}, {Hoskins}, {Hsieh}, {Jagannathan}, {Kawasaki}, {Keimpema}, {Kettenis}, {Lopez}, {Marvil}, {Masters}, {McNichols}, {Mehringer}, {Miel}, {Moellenbrock}, {Montesino}, {Nakazato}, {Ott}, {Petry}, {Pokorny}, {Raba}, {Rau}, {Schiebel}, {Schweighart}, {Sekhar}, {Shimada}, {Small}, {Steeb}, {Sugimoto}, {Suoranta}, {Tsutsumi}, {van Bemmel}, {Verkouter}, {Wells}, {Xiong}, {Szomoru}, {Griffith}, {Glendenning}, \& {Kern}}]{2022PASP..134k4501C}
{CASA Team}, {Bean}, B., {Bhatnagar}, S., {et~al.} 2022, \pasp, 134, 114501

\bibitem[{Cortes {et~al.}(2024)Cortes, Vlahakis, Hales, Carpenter, Dent, Kameno, Loomis, Vila-Vilaro, Immer, Law, Stoehr, \& Saini}]{almathbcy11}
Cortes, P.~C., Vlahakis, C., Hales, A., {et~al.} 2024, ALMA Technical Handbook Cycle 11 (ALMA Doc. 11.3, ver. 1.4), https://almascience.eso.org/documents-and-tools/cycle11/alma-technical-handbook

\bibitem[{{Darnley}(2021)}]{2021gacv.workE..44D}
{Darnley}, M.~J. 2021, in The Golden Age of Cataclysmic Variables and Related Objects V, Vol. 2-7, 44

\bibitem[{{de Ruiter} {et~al.}(2023){de Ruiter}, {Nyamai}, {Rowlinson}, {Wijers}, {O'Brien}, {Williams}, \& {Woudt}}]{2023MNRAS.523..132D}
{de Ruiter}, I., {Nyamai}, M.~M., {Rowlinson}, A., {et~al.} 2023, \mnras, 523, 132

\bibitem[{{Fekel} {et~al.}(2000){Fekel}, {Joyce}, {Hinkle}, \& {Skrutskie}}]{2000AJ....119.1375F}
{Fekel}, F.~C., {Joyce}, R.~R., {Hinkle}, K.~H., \& {Skrutskie}, M.~F. 2000, \aj, 119, 1375

\bibitem[{{Gaia Collaboration}(2020)}]{gaia_edr3}
{Gaia Collaboration}. 2020, Gaia EDR3 (CDS/ADC Collection of Electronic Catalogues), 1350, 0

\bibitem[{{Hinkle} {et~al.}(2025){Hinkle}, {Nagarajan}, {Fekel}, {Miko{\l}ajewska}, {Straniero}, \& {Muterspaugh}}]{2025ApJ...983...76H}
{Hinkle}, K.~H., {Nagarajan}, P., {Fekel}, F.~C., {et~al.} 2025, \apj, 983, 76

\bibitem[{{Hjellming} {et~al.}(1986){Hjellming}, {van Gorkom}, {Taylor}, {Sequist}, {Padin}, {Davis}, \& {Bode}}]{1986ApJ...305L..71H}
{Hjellming}, R.~M., {van Gorkom}, J.~H., {Taylor}, A.~R., {et~al.} 1986, ApJL, 305, L71

\bibitem[{{H{\"o}fner} \& {Olofsson}(2018)}]{2018A&ARv..26....1H}
{H{\"o}fner}, S. \& {Olofsson}, H. 2018, \aapr, 26, 1

\bibitem[{{Hunter} {et~al.}(2023){Hunter}, {Indebetouw}, {Brogan}, {Berry}, {Chang}, {Francke}, {Geers}, {G{\'o}mez}, {Hibbard}, {Humphreys}, {Kent}, {Kepley}, {Kunneriath}, {Lipnicky}, {Loomis}, {Mason}, {Masters}, {Maud}, {Muders}, {Sabater}, {Sugimoto}, {Sz{\H{u}}cs}, {Vasiliev}, {Videla}, {Villard}, {Williams}, {Xue}, \& {Yoon}}]{2023PASP..135g4501H}
{Hunter}, T.~R., {Indebetouw}, R., {Brogan}, C.~L., {et~al.} 2023, \pasp, 135, 074501

\bibitem[{{Jos{\'e}}(2016)}]{2016stex.book.....J}
{Jos{\'e}}, J. 2016, {Stellar Explosions: Hydrodynamics and Nucleosynthesis} (CRC Press)

\bibitem[{{Jos{\'e}} \& {Hernanz}(1998)}]{1998ApJ...494..680J}
{Jos{\'e}}, J. \& {Hernanz}, M. 1998, ApJ, 494, 680

\bibitem[{{Kato} \& {Hachisu}(2012)}]{2012BASI...40..393K}
{Kato}, M. \& {Hachisu}, I. 2012, Bulletin of the Astronomical Society of India, 40, 393

\bibitem[{Kloppenborg(2025)}]{aavso}
Kloppenborg, B. 2025, Observations from the AAVSO International Database, https://www.aavso.org

\bibitem[{{Linford} {et~al.}(2019){Linford}, {Chomiuk}, {Sokoloski}, {Weston}, {van der Horst}, {Mukai}, {Barrett}, {Mioduszewski}, \& {Rupen}}]{2019ApJ...884....8L}
{Linford}, J.~D., {Chomiuk}, L., {Sokoloski}, J.~L., {et~al.} 2019, \apj, 884, 8

\bibitem[{{L{\'o}pez-Coto} {et~al.}(2015){L{\'o}pez-Coto}, {Sitarek}, {Bednarek}, {de O{\~n}a Wilhelmi}, {MAGIC Collaboration}, {Desiante}, {Longo}, {Hays}, \& {Fermi-Lat Collaboration}}]{2015ICRC...34..731L}
{L{\'o}pez-Coto}, R., {Sitarek}, J., {Bednarek}, W., {et~al.} 2015, in International Cosmic Ray Conference, Vol.~34, 34th International Cosmic Ray Conference (ICRC2015), 731

\bibitem[{{Munari}(2023)}]{2023RNAAS...7..145M}
{Munari}, U. 2023, Research Notes of the American Astronomical Society, 7, 145

\bibitem[{{Munari}(2025)}]{2025CoSka..55c..47M}
{Munari}, U. 2025, Contributions of the Astronomical Observatory Skalnate Pleso, 55, 47

\bibitem[{{Munari} {et~al.}(2016){Munari}, {Dallaporta}, \& {Cherini}}]{2016NewA...47....7M}
{Munari}, U., {Dallaporta}, S., \& {Cherini}, G. 2016, \na, 47, 7

\bibitem[{{Munari} {et~al.}(2022){Munari}, {Giroletti}, {Marcote}, {O'Brien}, {Veres}, {Yang}, {Williams}, \& {Woudt}}]{2022A&A...666L...6M}
{Munari}, U., {Giroletti}, M., {Marcote}, B., {et~al.} 2022, \aap, 666, L6

\bibitem[{{M{\"u}rset} \& {Schmid}(1999)}]{1999A&AS..137..473M}
{M{\"u}rset}, U. \& {Schmid}, H.~M. 1999, \aaps, 137, 473

\bibitem[{{Nayana} {et~al.}(2024){Nayana}, {Anupama}, {Roy}, {Banerjee}, {Singh}, {Sonith}, \& {Kamath}}]{2024MNRAS.528.5528N}
{Nayana}, A.~J., {Anupama}, G.~C., {Roy}, N., {et~al.} 2024, MNRAS, 528, 5528

\bibitem[{{O'Brien} {et~al.}(2006){O'Brien}, {Bode}, {Porcas}, {Muxlow}, {Eyres}, {Beswick}, {Garrington}, {Davis}, \& {Evans}}]{2006Natur.442..279O}
{O'Brien}, T.~J., {Bode}, M.~F., {Porcas}, R.~W., {et~al.} 2006, Nature, 442, 279

\bibitem[{{Orlando} {et~al.}(2009){Orlando}, {Drake}, \& {Laming}}]{2009A&A...493.1049O}
{Orlando}, S., {Drake}, J.~J., \& {Laming}, J.~M. 2009, A\&A, 493, 1049

\bibitem[{{Panagia} \& {Felli}(1975)}]{1975A&A....39....1P}
{Panagia}, N. \& {Felli}, M. 1975, \aap, 39, 1

\bibitem[{{Planquart} {et~al.}(2025){Planquart}, {Jorissen}, \& {Van Winckel}}]{2025A&A...694A..85P}
{Planquart}, L., {Jorissen}, A., \& {Van Winckel}, H. 2025, \aap, 694, A85

\bibitem[{{Rau} \& {Cornwell}(2011)}]{2011A&A...532A..71R}
{Rau}, U. \& {Cornwell}, T.~J. 2011, \aap, 532, A71

\bibitem[{{Rupen} {et~al.}(2008){Rupen}, {Mioduszewski}, \& {Sokoloski}}]{2008ApJ...688..559R}
{Rupen}, M.~P., {Mioduszewski}, A.~J., \& {Sokoloski}, J.~L. 2008, ApJ, 688, 559

\bibitem[{{Schaefer}(2023{\natexlab{a}})}]{2023MNRAS.524.3146S}
{Schaefer}, B.~E. 2023{\natexlab{a}}, \mnras, 524, 3146

\bibitem[{{Schaefer}(2023{\natexlab{b}})}]{2023JHA....54..436S}
{Schaefer}, B.~E. 2023{\natexlab{b}}, Journal for the History of Astronomy, 54, 436

\bibitem[{{Shafter}(2017)}]{2017ApJ...834..196S}
{Shafter}, A.~W. 2017, ApJ, 834, 196

\bibitem[{{Shore}(2008)}]{2008ASPC..401...19S}
{Shore}, S.~N. 2008, in Astronomical Society of the Pacific Conference Series, Vol. 401, RS Ophiuchi (2006) and the Recurrent Nova Phenomenon, ed. A.~{Evans}, M.~F. {Bode}, T.~J. {O'Brien}, \& M.~J. {Darnley}, 19

\bibitem[{{Shore} \& {Aufdenberg}(1993)}]{1993ApJ...416..355S}
{Shore}, S.~N. \& {Aufdenberg}, J.~P. 1993, ApJ, 416, 355

\bibitem[{{Sokoloski} {et~al.}(2006){Sokoloski}, {Luna}, {Mukai}, \& {Kenyon}}]{2006Natur.442..276S}
{Sokoloski}, J.~L., {Luna}, G.~J.~M., {Mukai}, K., \& {Kenyon}, S.~J. 2006, Nature, 442, 276

\bibitem[{{Sokoloski} {et~al.}(2008){Sokoloski}, {Rupen}, \& {Mioduszewski}}]{2008ApJ...685L.137S}
{Sokoloski}, J.~L., {Rupen}, M.~P., \& {Mioduszewski}, A.~J. 2008, ApJL, 685, L137

\bibitem[{{Stanishev} {et~al.}(2004){Stanishev}, {Zamanov}, {Tomov}, \& {Marziani}}]{2004A&A...415..609S}
{Stanishev}, V., {Zamanov}, R., {Tomov}, N., \& {Marziani}, P. 2004, \aap, 415, 609

\bibitem[{{Starrfield} {et~al.}(2025){Starrfield}, {Bose}, {Woodward}, {Iliadis}, {Hix}, {Evans}, {Shaw}, {Banerjee}, {Liimets}, {Page}, {Geballe}, {Ilyin}, {Perron}, \& {Wagner}}]{2025ApJ...982...89S}
{Starrfield}, S., {Bose}, M., {Woodward}, C.~E., {et~al.} 2025, \apj, 982, 89

\bibitem[{{Starrfield} {et~al.}(2016){Starrfield}, {Iliadis}, \& {Hix}}]{2016PASP..128e1001S}
{Starrfield}, S., {Iliadis}, C., \& {Hix}, W.~R. 2016, PASP, 128, 051001

\bibitem[{{Straizys} \& {Kuriliene}(1981)}]{1981Ap&SS..80..353S}
{Straizys}, V. \& {Kuriliene}, G. 1981, \apss, 80, 353

\bibitem[{{Taylor} \& {Seaquist}(1984)}]{1984ApJ...286..263T}
{Taylor}, A.~R. \& {Seaquist}, E.~R. 1984, \apj, 286, 263

\bibitem[{{Townsley}(2008)}]{2008ASPC..401..131T}
{Townsley}, D.~M. 2008, in Astronomical Society of the Pacific Conference Series, Vol. 401, RS Ophiuchi (2006) and the Recurrent Nova Phenomenon, ed. A.~{Evans}, M.~F. {Bode}, T.~J. {O'Brien}, \& M.~J. {Darnley}, 131

\bibitem[{{van Hoof} {et~al.}(2014){van Hoof}, {Williams}, {Volk}, {Chatzikos}, {Ferland}, {Lykins}, {Porter}, \& {Wang}}]{2014MNRAS.444..420V}
{van Hoof}, P.~A.~M., {Williams}, R.~J.~R., {Volk}, K., {et~al.} 2014, \mnras, 444, 420

\bibitem[{{Walder} {et~al.}(2008){Walder}, {Folini}, \& {Shore}}]{2008A&A...484L...9W}
{Walder}, R., {Folini}, D., \& {Shore}, S.~N. 2008, A\&A, 484, L9

\bibitem[{{Wright} \& {Barlow}(1975)}]{1975MNRAS.170...41W}
{Wright}, A.~E. \& {Barlow}, M.~J. 1975, \mnras, 170, 41

\bibitem[{{Yudin} \& {Munari}(1993)}]{1993A&A...270..165Y}
{Yudin}, B. \& {Munari}, U. 1993, \aap, 270, 165

\bibitem[{{Zamanov} {et~al.}(2023){Zamanov}, {Boeva}, {Latev}, {Semkov}, {Minev}, {Kostov}, {Bode}, {Marchev}, \& {Marchev}}]{2023A&A...680L..18Z}
{Zamanov}, R., {Boeva}, S., {Latev}, G.~Y., {et~al.} 2023, \aap, 680, L18

\bibitem[{{Zamanov} {et~al.}(2024){Zamanov}, {Stoyanov}, {Marchev}, {Minev}, {Marchev}, {Moyseev}, {Mart{\'\i}}, {Bode}, {Konstantinova-Antova}, \& {Stefanov}}]{2024AN....34540036Z}
{Zamanov}, R.~K., {Stoyanov}, K.~A., {Marchev}, V., {et~al.} 2024, Astronomische Nachrichten, 345, e20240036

\end{thebibliography}

\end{document}